\documentclass[final,twocolumn,oneside,10pt,a4paper]{IEEEtran} %,compsoc
\usepackage{times,subfigure}
\usepackage{graphicx,algorithm,algorithmic,multirow,flushend}
\usepackage{pgf,tikz}
\usepackage{graphicx,multirow}
\usepackage{latexsym,color,float,amsmath,amssymb,amsthm}
\theoremstyle{definition}

\theoremstyle{plain}

\theoremstyle{remark}

\newcounter{tmp-counter}

\begin{document}

\title{Link Scheduling in Amplify-and-Forward Cooperative Wireless Networks}%

% and Dimitrios Kosmanos,~\IEEEmembership{Student Member,~IEEE}
\author{Antonios Argyriou,~\IEEEmembership{Member,~IEEE}\thanks{The author is with The Department of Electrical \& Computer Engineering, University of Thessaly, Volos, 38221, Greece.}\thanks{Antonios Argyriou would like to acknowledge the support from the European Commission through the Marie Curie Intra-European Fellowship WINIE-273041.}% and the STREP project CONECT (FP7ICT257616).}
}

%\IEEEauthorblockA{\IEEEauthorrefmark{1}Department of Computer and Communication Engineering, University of Thessaly, Volos, 38221, Greece}%
%\IEEEauthorblockA{\IEEEauthorrefmark{2}Philips Research, Eindhoven, 5656AE, The Netherlands }
%\IEEEauthorblockA{\IEEEauthorrefmark{3}Eurandom, TU/e, Eindhoven, 5656AE, The Netherlands }

%}

%\IEEEpubid{0000--0000/00\$00.00~\copyright~200X IEEE}

\maketitle%

%\graphicspath{{figs3/}}

\markboth{}{Link Scheduling in Amplify-and-Forward Cooperative Wireless Networks}

%Important paper and the modeling approach for relay search algorithms is ~\cite{erkip05}

\begin{abstract}
In this letter we are concerned with the problem of link scheduling for throughput maximization in wireless networks that employ a cooperative amplify and forward (AF) protocol. To address this problem first we define the signal-to-interference plus noise ratio (SINR) expression for the complete cooperative AF-based transmission. Next, we formulate the problem of link scheduling as a mixed integer linear program (MILP) that uses as a constraint the developed SINR expression. The proposed formulation is motivated by the observation that the aggregate interference that affects a single cooperative transmission can be decomposed into two separate SINR constraints. Results for the optimal solution and a polynomial time approximation algorithm are also presented.
%Based on the developed interference model, and by exploiting the behavior of the AF protocol, the problem is formulated in such a way that it allows the decoupling of the signal-to-interference plus noise ratio (SINR) for the direct and cooperative transmissions.
%Numerical results are obtained for different traffic loads, number of relays.
\end{abstract}


\section{Introduction}
Link scheduling in a wireless network consists of the activation of point-to-point links between source-destination pairs at specific time instants~\cite{kumaran03}. A schedule that is optimized will allow more network nodes to transmit concurrently by minimizing interference to each other. %For link scheduling to take place, there must be a way to quantify precisely the interfering relationships between network nodes. The protocol and the interference models have been used extensively for this purpose while the later captures more accurately the physical process of wireless transmissions~\cite{gupta00}.
Existing link scheduling algorithms that model the transmission of wireless signals with the interference model~\cite{gupta00}, perceive the network as a collection of single transmitter-single receiver point-to-point links.

\begin{figure}[t]
\begin{center}
  \includegraphics[scale=1]{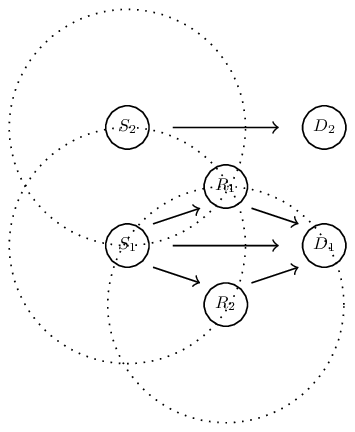}
  \includegraphics[scale=1]{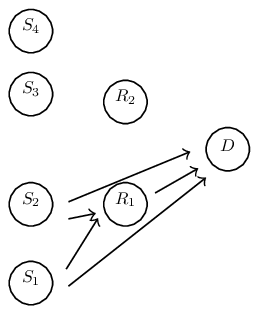}
 \end{center}
 %%\input{figure1.tex}
%%\input{figure2.tex}
%}
\caption{Wireless network scenarios that use amplify-and-forward at the cooperative relays. The dashed lines depict the transmission range of the respective nodes. %In (a) if node $S_2$ is scheduled simultaneously with $S_1$ the optimal relay may be $R_2$ and not $R_1$. In (b) the relay may forward the signal from the source that can achieve the maximum throughput improvement.
}
\label{fig:ls3-coop-topology}
\end{figure}

%However, in emerging wireless networks, not all transmissions can be captured with the previous model. In the recent years there is a shift towards changing the point-to-point paradigm with the aid of helping nodes, usually named relays~\cite{meulen68}. A relay in a wireless network essentially alters the point-to-point communication link between two nodes since it involves a third node. In this case the forwarding of a signal from the relay is an integral part of the primary transmission of the initial source since the destination node jointly decodes the two signals~\cite{laneman01}. This behavior is completely different from cooperation that requires decoding at the relays.
However, when node cooperation is employed through relays~\cite{laneman01}, the concept of a communication link is altered since a third node is involved. In this case link scheduling becomes a tougher problem to address. The two aspects of the same fundamental problem that arises in this case can be explained with the network topology depicted in Fig.~\ref{fig:ls3-coop-topology}(a). When source node $S_1$ selects node $R_1$ to be the cooperative relay that amplifies and forwards (AF) its signal destined to $D_1$, the interference generated by this "composite" transmission consists of the interference generated from both $S_1$ and $R_1$ (although in different time slots). The second problem is that this cooperative transmission that originates from $S_1$ is affected by the aggregate interference that is accumulated at node $D_1$ and also at relay $R_1$ (another node $S_2$ interferes in this case). The same type of problems can also occur in networks that employ a centralized architecture (Fig.~\ref{fig:ls3-coop-topology}(b)). %In this case relays may be employed to aid in uplink or downlink transmissions as the example in Fig.~\ref{fig:ls3-coop-topology}(b) indicates. Even with the presence of a single relay and single destination, the same problem occurs.

%\section{Related Works}
Link scheduling that considers point-to-point wireless links has been studied extensively in the literature. However, the impact of relay nodes that AF signals has not attracted significant attention. Hong~\emph{et. al} in~\cite{hong09,hong10} considered relays as a mechanism to aid in multi-hop communication but without using them as a means to improve the PHY performance. The problem is casted as a minimization of the total interference across multiple hops. % The same authors considered a simpler form of the same problem but with the joint objective of routing and scheduling in~\cite{hong09}.
Recently, Goussevskaia and Wattenhofer analyzed the complexity of scheduling wireless links under the physical interference model with a single relay~\cite{olga08}. The authors considered a canonical network with a specific line structure and bidirectional traffic flow. Xue~\emph{et. al} in~\cite{xue07a} considered opportunistic scheduling for two-way physical layer network coding in line networks with bidirectional traffic.

In this letter we investigate link scheduling for a wireless network that employs relays that use an AF protocol. To address the problem of link scheduling in this setting, first we define an extended physical interference model for AF-based cooperative transmissions. Next, we decouple the signal-to-interference plus noise ratio (SINR) expression of the cooperative transmission and we proceed with the formulation of the link scheduling throughput maximization problem as a mixed integer linear program (MILP).

\section{System Model}
\label{section:system-model}
The network model we define in this letter consists of a set $\mathcal{S}$ of $N$ source nodes that want to communicate with a group of $D$ destination nodes denoted as the set $\mathcal{D}$. Communication can be accomplished with the assistance of $M$ relays that belong in the set $\mathcal{R}$. Each destination may also have multiple incoming sources. Thus, the network model is generic and can reflect either distributed or centralized network topologies (Fig.~\ref{fig:ls3-coop-topology}). To accommodate the previous design choice we denote the existence of a link from a source $i$ to a destination $j$ as $l_{ij}$, while all these links are contained in the set $\mathcal{L}$. %The link from relay $r$ to a destination node $j$ is denoted as $l_{rj}$.
Since a particular relay might serve any of the valid source/destination links $l_{ij}$, there is a need to define the extended group of point-to-point relay links. This group $\mathcal{R}^{'}\triangleq \{l_{irj}, \forall (i,r,j)\in \mathcal{S}\times\mathcal{R}\times\mathcal{D}\}$ consists of all the combinations of source/relay/destination links.

%The complete network is modeled with a directed graph $F(\mathcal{U},\mathcal{V})$, where $\mathcal{U}$ and $\mathcal{V}$ are the set of point-to-point directional links, and the set of nodes, respectively ($\mathcal{V} = \{\mathcal{S},\mathcal{D},\mathcal{R}^{'}\}$). Therefore, the number of point-to-point links in the graph is $N$ if direct transmission is used. Each vertex in the conflict graph represents a wireless link in the network, and there is an edge between two vertices if and only if the links represented by the vertices conflict (i.e. they interfere with each other and simultaneous transmission is impossible). On the other hand, a clique in the conflict graph represents a group of links that cannot transmit concurrently, and hence they must access the channel exclusively.

\begin{figure}[t]%p to throw everything at the end
 \begin{center}
  \includegraphics[scale=1]{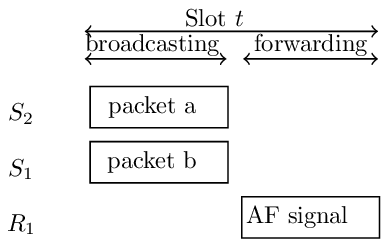}\\
 \end{center}
\caption{Behavior of the cooperative protocol in the time domain. Time is slotted and each slot is separated in two phases.}
\label{fig:ls3-coop-protocol}
\end{figure}

%\subsection{System Model Cooperative Protocol}
The basic cooperative protocol we present adopts two distinct phases and its messaging structure can be seen in Fig.~\ref{fig:ls3-coop-protocol}. During the broadcasting phase, the source nodes broadcast depending on whether they were scheduled or not, and these signals are received by all nodes at different power levels (including the relays). During the forwarding phase the relays do not decode the packets but instead they AF the received signal if they are activated~\cite{laneman04}. The sources and relays that are active in slot $t$ are denoted as $\mathcal{S}_t$, and $\mathcal{R}^{'}_t$ respectively.

\subsection{Interference Model}
Let us now define the interference model for the proposed system. %The aggregate interference that is accumulated over a node $i$ during slot $t$ depends on the transmission phase.
During broadcasting all the sources that transmit will interfere with each other. For the particular link $l_{ij}$ the power of the aggregate interference is expressed as (the first subscript denotes the destination and the second the phase)
\begin{equation}\label{eqn:I_jb}
I_{j,b}= \sum_{l_{mk}\in \mathcal{S}_t - \{l_{ij}\}} \gamma_{mj} P^t_m,
\end{equation}
where $l_{mk}$ is an auxiliary variable that counts all the links that are active in slot $t$. $P^t_m$ corresponds to the transmitter power of a source $m$ during slot $t$. Also $\gamma_{mj}=1/{d^a_{mj}}$, where $d_{mj}$ is the distance between source $m$ and destination $j$ while $a$ is usually set to a value between 3 and 4. During the same broadcasting phase, interference will also be present at relay nodes. The aggregate power in this case that is denoted as $I_{r,b}$ and is captured with the same formula given in~\eqref{eqn:I_jb}.

During forwarding, interference will be generated from the relays that forward their signals. This can expressed for destination $j$ as:
\begin{equation}\label{eqn:I_jf}
I_{j,f}= \sum_{l_{mk}\in \mathcal{R}'_t - \{l_{irj}\}} \gamma_{mj} P^t_m
\end{equation}
Now we can define the SINR of the cooperative transmission that occurs in two orthogonal time slots as
\begin{eqnarray}
\label{eqn:snr_coop}
SINR_{i,j,r}=\frac{P^t_i\gamma_{ij}}{I_{j,b}+\sigma^2}+\frac{P^t_i\gamma_{ir}\gamma_{rj}g_r^2}{\sigma^2+I_{j,f}+(\sigma^2+I_{r,b})\gamma_{rj}g_r^2} .
\end{eqnarray}
In the above $\sigma^2$ is the AWGN variance. Note that the previous expression is an extension of the basic cooperative diversity formula~\cite{laneman04}. In the above expression the relay scales the received signal by a factor $g_r$ so as to maintain the power constraint~\cite{laneman04}.
%\begin{eqnarray}\label{eqn:power-constraint1}
%g_r =  \sqrt{\frac{P^t_i}{P^t_i \gamma_{ir} +\sigma^2}}.
%\end{eqnarray}

\begin{figure*}[t]
\begin{eqnarray*}\label{eqn:problem5}
\text{max.} &&  \frac{1}{T} \sum_{t=1}^T \sum_{l_{ij}\in \mathcal{L}} x^{t}_{ij} \\%+ \sum_{r=1}^{M} \sum_{j=1}^{D} \sum_{i=1}^{N}  y^t_{irj}) \quad \\
\text{s. t.}  &&   x^{t}_{ij},y^{t}_{irj} \in \{ 0,1 \} ~ (C1),~ \sum^{T}_{t=1} x^{t}_{ij} \geq B_{i},  \forall l_{ij}\in \mathcal{L}~(C2),~ \frac{P^t_{i}\gamma_{ij}+(1-x^{t}_{ij})\Delta}{\sigma^2+\sum_{k\in\mathcal{S}_t-\{i\}} P^t_{k}\gamma_{kj}} \geq \beta_1+(1-y^{t}_{irj})\beta_2, ~ \forall l_{ij}\in \mathcal{L}, \forall t\in T ~ (C3)\\
&& \frac{P^t_{i}\gamma_{ir}\gamma_{rj}g_r^2+(1-y^{t}_{irj})\Delta}{\sigma^2+\sum_{k\in\mathcal{R}'-\{r\}} P^t_{k}\gamma_{kj}+(\sigma^2+\sum_{k\in\mathcal{S}_t-\{i\}} P^t_{k}\gamma_{kr})\gamma_{ri}g_r^2}\geq \beta_2,  ~ \forall l_{irj}\in \mathcal{R}', \forall t\in T ~ (C4),~\sum_{r=1}^{M}y^t_{irj}\leq 1~\forall i\in \mathcal{S}, \forall t\in T  (C5)\\
&&  x^{t}_{ij} \geq y^{t}_{irj}, \forall i\in \mathcal{S},\forall l_{irj}\in \mathcal{R}'~ (C6),~ \sum^{T}_{t=1} P^{t}_{i} \leq P^{max}_{i}~ (C7),  ~ x^{t}_{ij} P^{t,min}_{i}\leq P^{t}_i \leq x^{t}_{ij} P^{t,max}_{i} ~ (C8).\\
\hline
\end{eqnarray*}
\end{figure*}

\section{MILP for Throughput Maximization in Wireless Relay Networks}
The objective of the $\mathbf{CLS}$~(cooperative link scheduling) problem formulation is to maximize the throughput in an arbitrary wireless network by allowing the dynamic activation of the links that originate from  sources and relays. To model this link scheduling problem, in the formulation we introduce binary optimization variables that indicate whether a link is active or not. First, we define variable $x^t_{ij}$ that denotes whether source $i$ transmits to its destination $j$ during slot $t$. Next, we define variable $y^t_{irj}$ that indicates whether the point-to-point links from source $i$ to relay $r$ and from relay $r$ to destination $j$ are activated or not. We also introduce the continuous optimization variable $P^t_i$ that corresponds to the transmission power of source $i$. Thus, the optimization problem constitutes a mixed integer linear program. We also denote with $T$ the maximum number of slots for which the problem is solved, and finally $\beta$ is the SINR decoding threshold at the destination.

\emph{The most important feature of the problem formulation is that it separates the SINR expression of the cooperative transmission so that the interference constraints maintain a linear form. The key observation that allows the above approach is that the precise SINR values of the broadcasting and forwarding transmissions for a specific packet and a specific destination do not matter as long as their sum is higher or equal to $\beta$, i.e. if the packet is decodable at the final destination after forwarding.} To implement the previous insight in practice we separate the cooperative SINR expression shown in~\eqref{eqn:snr_coop} into the direct and AF transmission parts by introducing the \emph{auxiliary packet decoding thresholds} $\beta_1$ and $\beta_2$. The detailed problem formulation is explained below.

% This practically means if the condition that must hold is $SINR_{DIR}+ SINR_{COOP}\geq \beta$.

The objective is to maximize the throughput by increasing the number of scheduled valid source-destination pairs from the set $\mathcal{L}$. The first constraint, C1 refers to the integer optimization variables. Next, C2 ensures that each source $i$ is scheduled at least $B_{i}$ slots which may be an optimal constraint. C3 refers to the SINR of the direct transmission that must be higher than the required threshold $\beta_1$. $\Delta$ is a high value constant that essentially disables this constraint when the link is not scheduled ($x^t_{ij}=0$). With our formulation a direct transmission can be scheduled if the resulting SINR is higher than $\beta_1$ which is not necessarily the same with $\beta$. This means that the scheduling condition for link $l_{ij}$ can be less restrictive. C4 that is slightly more complicated, and it corresponds to the SINR of the AF transmission that must be at least equal to the second threshold $\beta_2$. In C4 the interference summations that are included are first the collective interference during the relay forwarding phase, and also the interference from the sources that were activated when source $i$ was broadcasting. For the packet to be decoded, the value of the aggregate SINR must be higher than $\beta$ and so we must set $\beta_1+\beta_2$ equal to $\beta$. C5 ensures that a source $i$ that transmits in slot $t$ will either use one relay or not. C6 is formulated in a very simple form but it is critical since it ensures that a relay link cannot be active before the corresponding source link was activated. With this constraint it is possible that a source was activated in slot $t$ and the relay was not activated in that slot, but not the other way around. %Although a signal from a transmission might be received at every network node, note that with the term activation we mean the selection of a proper power level so  that the corresponding destination can decode the packet (i.e. the SINR is above $\beta$).
C7 ensures that the total power expenditure for a source node during the $T$ slots is within a certain budget. The source must comply with this power budget regardless of how many times it was activated. Finally, C8 ensures the per-slot transmitter power constraint. %transceiver.

\begin{figure}[t]
\framebox[1.0\linewidth]{
\begin{minipage}[t]{0.9\linewidth}
$randomized\_rounding\_contraint\_check()$
\begin{algorithmic}[1]
%\STATE $\hat{\mathbf{x}},\hat{\mathbf{P}}=lp(\mathcal{S},\mathcal{D},T,B)$ //find the solution vectors
%\FOR{all nodes $i\in \mathcal{S}$ }
%\STATE $c_i=\sum^{D_i}_{j=1}\sum^{T}_{t=1} \hat{x}^{t}_{ij}$
%\ENDFOR
%\STATE $\mathcal{Z}=\mathcal{S}$
\FOR{all links }
\STATE Assign $\hat{x}^t_{ij},\hat{y}^t_{irj}$ according to~\eqref{eqn:randomized-rounding}
%\WHILE{$\hat{x}_{ij}$ not feasible}
\IF{$\hat{x}^t_{ij}=1$ AND $C1=true$}
\STATE continue; //C6 is valid anyway
\ELSIF{$\hat{x}^t_{ij}=1$ AND $C1!=true$}
\STATE $\hat{x}^t_{ij}\leftarrow 0$; //C1 will become valid because of $\Delta$
\ENDIF
\IF{$\hat{x}^t_{ij}=0$ AND $P^t_i !=0$ }
\STATE set  $\hat{P}^t_i \leftarrow 0$; //So that C7 becomes valid
\ELSE
\STATE continue //C1 will be valid anyway
\ENDIF
%\ENDWHILE ~~// Solution for $i$: $\tilde{\mathbf{x}}_{i},\tilde{\mathbf{P}}_{i}$
\IF{$\hat{y}^t_{irj}=1$ AND $\hat{x}^t_{ij}=1$  }
\IF{$C2!=true$}
\STATE $\hat{y}^t_{irj}\leftarrow 0$ ;
\ENDIF
\ELSIF{$\hat{y}^t_{irj}=1$ AND $\hat{x}^t_{ij}=0$  }
\STATE set  $\hat{y}^t_{irj}\leftarrow 0$;
\ELSIF{$\hat{y}^t_{ijr}=0$  }
\STATE continue;
\ENDIF
\ENDFOR
\end{algorithmic}
\end{minipage}
}
\caption{Pseudo-code for checking the constraints after applying randomized rounding to the LP solution.}%
\label{fig:scheduling-algorithm}
\end{figure}

\subsection{MILP Relaxation and Approximation Algorithm}
Since there are no polynomial time algorithms for solving MILPs, we relax the original problem so that a linear program (LP) can be solved. LPs can be solved in polynomial time with interior point methods. In this particular case we allow the indicator variables $x^t_{ij},y^t_{irj}$ to take any value between 0 and 1. After the LP is solved, the result of the relaxed LP that consists of a set of continuous values between 0 and 1 must be converted to a binary value of either 0 or 1. Thus, a suboptimal and approximate approach has to be designed for performing the previously mentioned assignments. To this aim, we adopt the well known randomized rounding algorithm for creating our heuristic~\cite{randomize-rounding,fan09b}. The main idea of the approximation algorithm to assign the final binary values with a certain probability as follows. If we denote with $\tilde{x}^t_{ij},\tilde{y}^t_{ijr},\tilde{P}^t_i$ the solutions of the LP, the binary values are approximated with randomized rounding as follows:
 \begin{eqnarray}\label{eqn:randomized-rounding}
 \hat{x}^t_{ij} =\left \{ \begin{array}{ll}
   P_r[1]= \tilde{x}^t_{ij}   \\
   P_r[0]=1-\tilde{x}^t_{ij}
  \end{array}\right.
   \hat{y}^t_{ijr} =\left \{ \begin{array}{ll}
  P_r[1]= \tilde{y}^t_{irj}   \\
  P_r[0]=1-\tilde{y}^t_{irj}
  \end{array}\right.
  \end{eqnarray}
The above rule means that the final binary solution $\hat{x}^t_{ij}$ is equal to 1 w.p. $\tilde{x}^t_{ij}$ and equal to 0 w.p. $1-\tilde{x}^t_{ij}$. A value for $\tilde{x}^t_{ij}$ closer to 1 increases the probability that a binary 1 is actually assigned. The above process ensures that the expectation of the cost of the MILP and LP solutions are the same~\cite{randomize-rounding}. Since some constraints might be invalid after the assignment in~\eqref{eqn:randomized-rounding}, they must be verified before we obtain the final result. This is accomplished with the algorithm depicted in Fig~\ref{fig:scheduling-algorithm}. Consider for example the case that $\hat{x}_{ij}=1$. It is possible that C3 does not hold since the solution of the LP can have an optimal value $\tilde{x}_{ij}$ lower than 1, that could lead to an arbitrarily low power level $\tilde{P}^t_i$. If this is the case, our approximation algorithm does not schedule the link by setting $\hat{x}_{ij}\leftarrow 0$ and $\hat{P}^t_i\leftarrow 0$. For the variable $y^t_{irj}$ the algorithm must only check if the corresponding source was activated during broadcasting (line 13). Similar reasoning follows for the remaining cases.% that are not explained in detail due to lack of space.

\subsection{Impact of Packet Buffers}
A note should be made here regarding the ability of the proposed approach to take into account the number of buffered packets at each node. In this case the network throughput should be maximized for multiple periods each of duration $T$. The buffered packets should be taken into account and a priority should be given to nodes that have more buffered packets. This could be accomplished by changing $B_i$ for a specific node every time the problem is solved. One potential metric would be to assign $B_i=\lfloor \frac{\text{Buffered packets at }i}{\text{Buffer size at }i}T \rfloor$, where sources with higher buffer occupancy are prioritized. This approach could also be used for implementing different fairness policies and for adaptation to dynamic traffic conditions.

%\subsection{Applications/Simplifications of the Basic Model}
%\textbf{Uplink/Downlink Centralized Wireless Network:} In the cellular uplink scheduling scenario $\mathcal{D}$ consists of one node. Also if a single relay serves a group of nodes the same condition will hold for the group of relays.

\section{Performance Evaluation}
\label{section:performance-evaluation}
%\subsection{Setup}
In this section we compare the performance of $\mathbf{CLS}$ with that of direct link scheduling ($\mathbf{DLS}$). The purpose of the comparison is to help identify the relative performance benefits of the cooperative approach. Most of the results correspond to the optimal solution calculated with CPLEX Optimization Studio V12.5.0 while we also have results for the approximation algorithm. For $\mathbf{CLS}$ we test different numbers of maximum available relays. The system parameters are set as follows: $\beta=10$dB,  $P^{t,max}_i,g^2_r$=300mW, $P^{t,min}_i$=0.01$P^{t,max}_i$, $P$=0.3$P^{t,max}_i$. Node distances $d^a_{i,j}$ are randomly and uniformly selected in the range $[0,100]$ with $a$=3. %We experimented with different traffic loads in terms of the required slots that must be active $B$, but in each experiment all the nodes were configured with the same load.
In the figures the horizontal axis depicts the number of sources $N$ and the vertical axis the normalized throughput. Throughput is calculated by considering all the successful transmissions from all the sources.% (transmissions beyond the threshold $\beta$).

\begin{figure}[t]
\begin{center}
\subfigure[$\beta_1=\beta_2=5$dB]{\includegraphics[keepaspectratio,width = 0.49\linewidth]{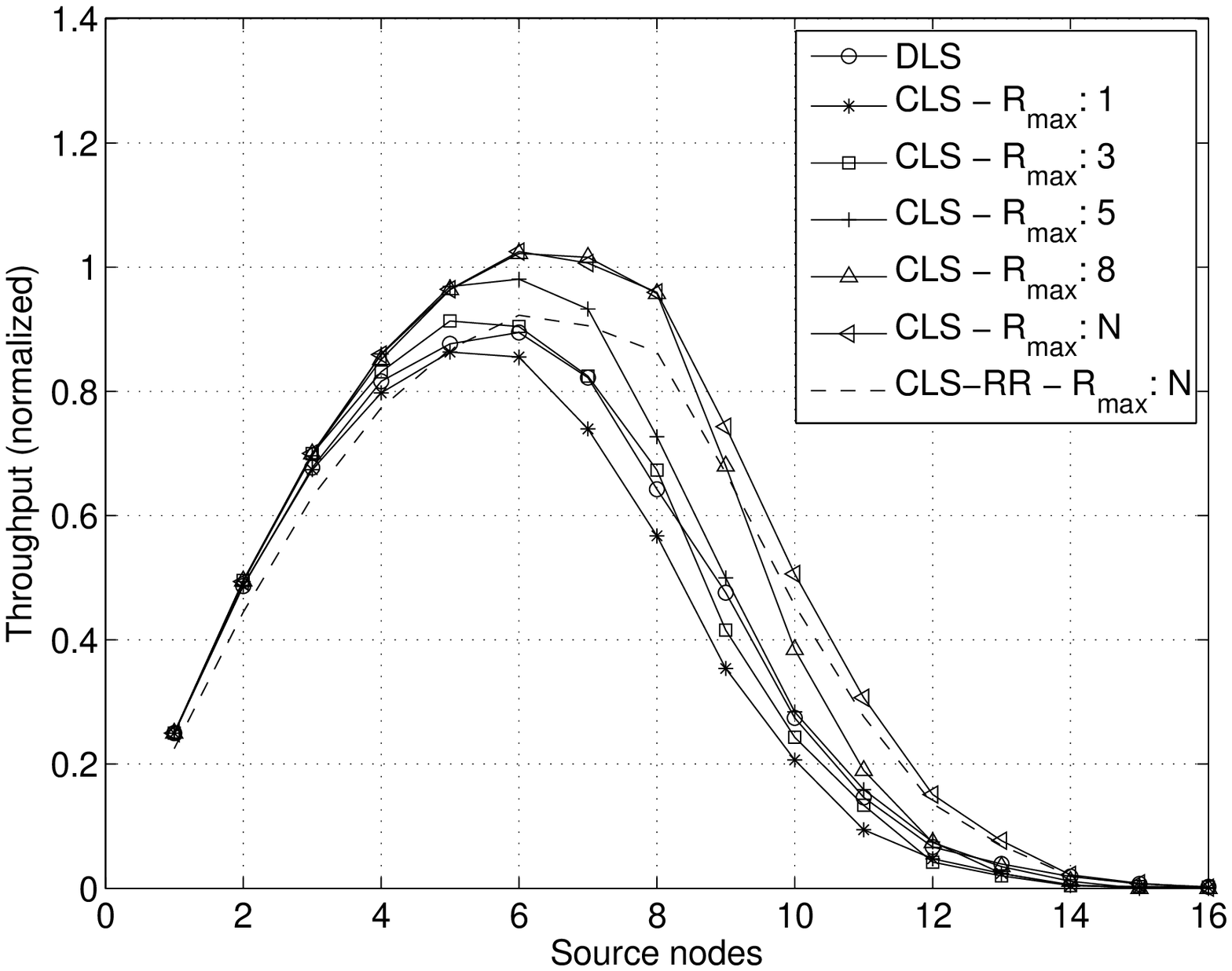}}
\subfigure[$\beta_1=4$dB,$\beta_2=6$dB]{\includegraphics[keepaspectratio,width = 0.49\linewidth]{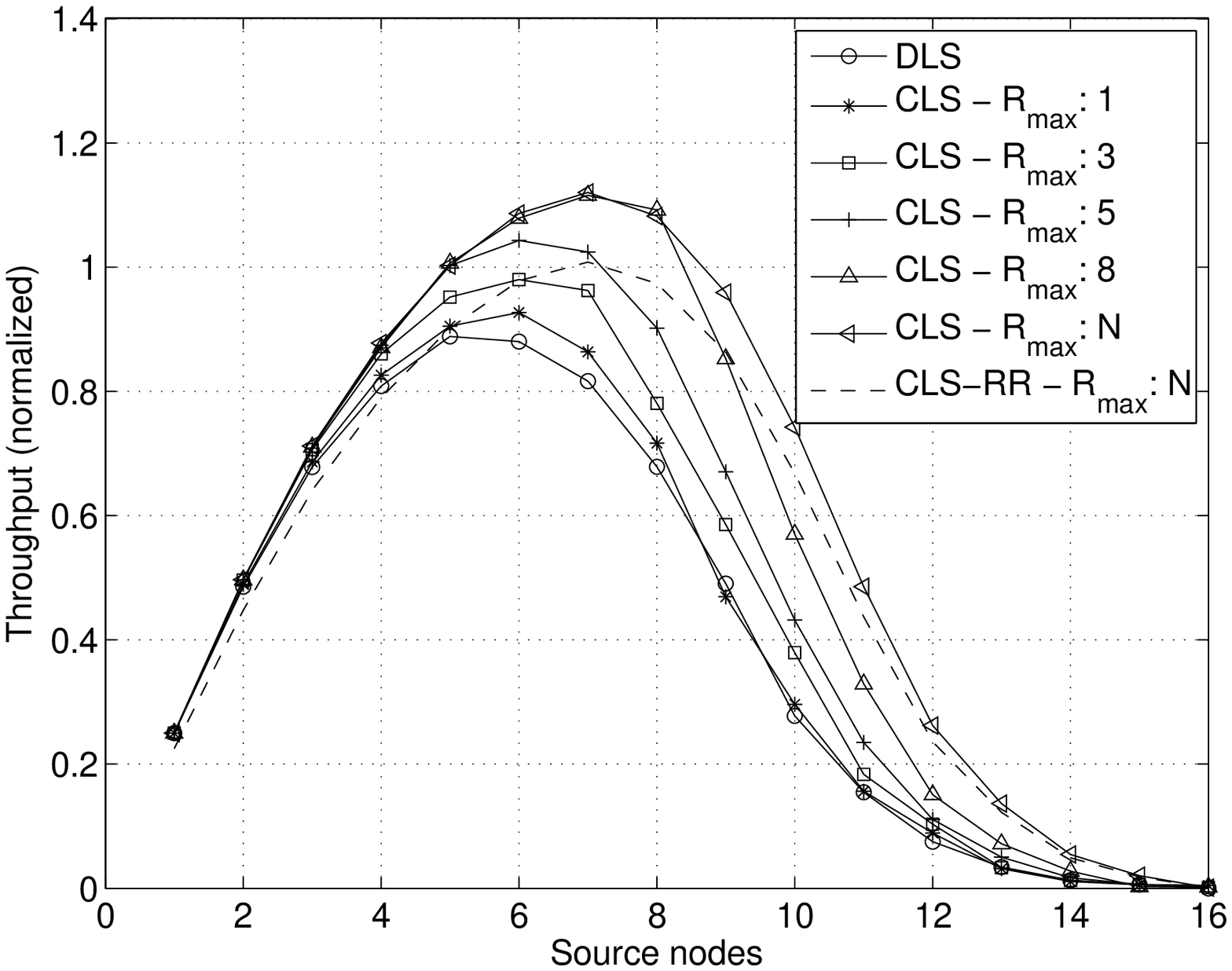}}
 \end{center}
 \caption{$B$=$T$=8 slots.}
 \label{fig:results1}
\end{figure}

\begin{figure}[t]
\begin{center}
\subfigure[$\beta_1=\beta_2=5$dB]{\includegraphics[keepaspectratio,width = 0.49\linewidth]{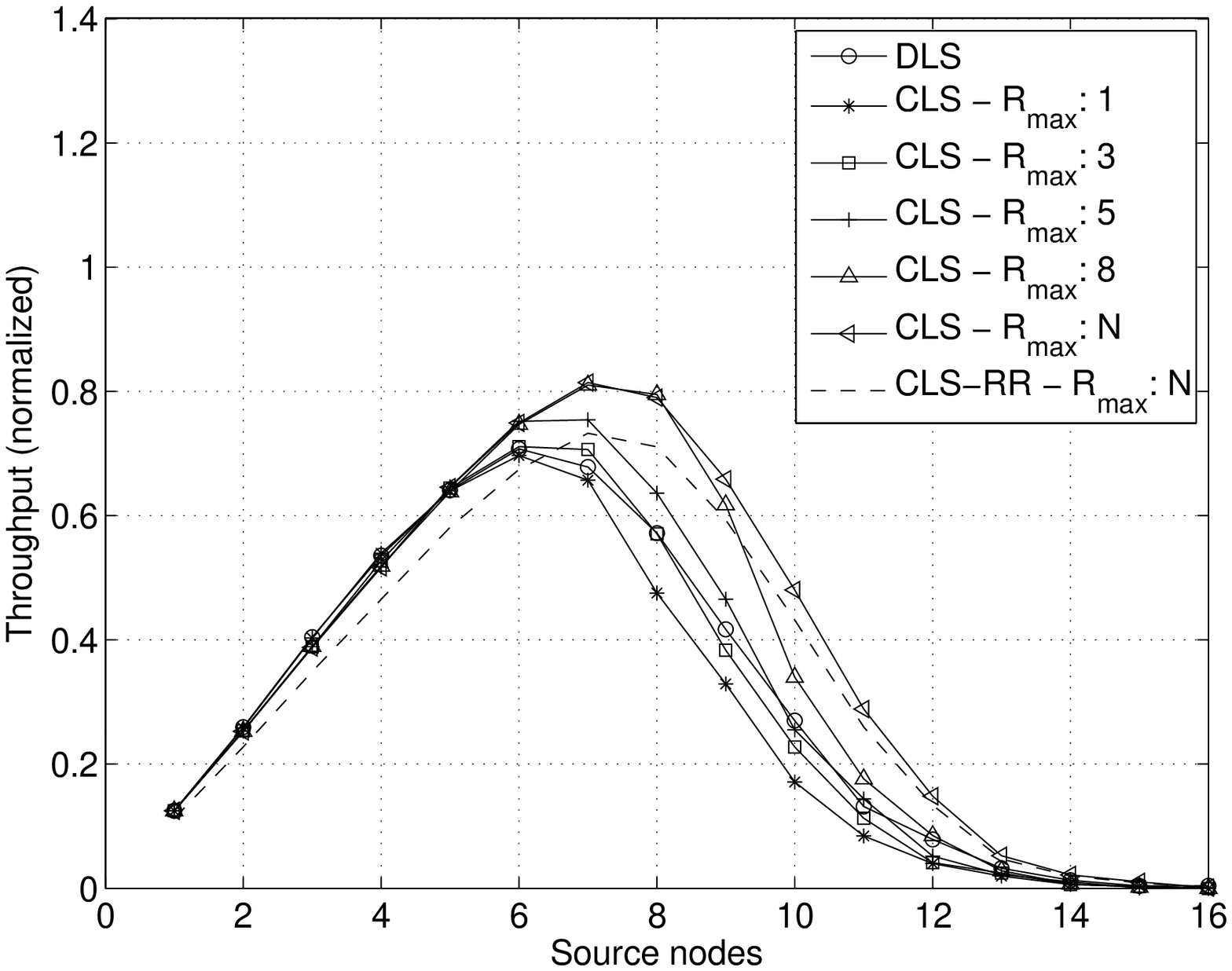}}
\subfigure[$\beta_1=4$dB,$\beta_2=6$dB]{\includegraphics[keepaspectratio,width = 0.49\linewidth]{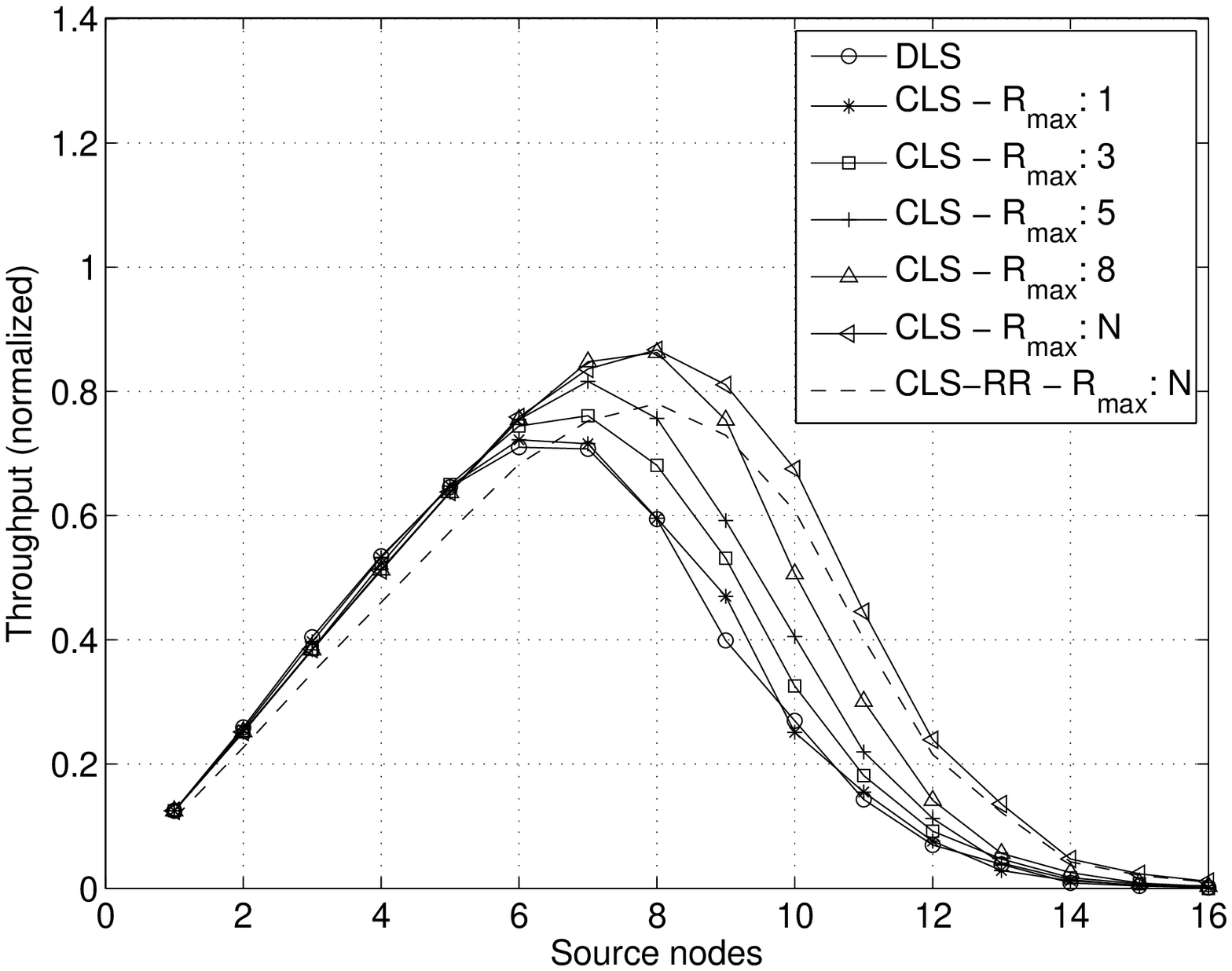}}
 \end{center}
 \caption{$B$=4,$T$=8 slots.}
 \label{fig:results2}
\end{figure}

In Fig.~\ref{fig:results1} the performance is presented both for $\mathbf{CLS}$ and $\mathbf{DLS}$. In Fig.~\ref{fig:results1} $B$=$T$=8 which means that all the nodes are backlogged and desire to transmit a data packet in every slot. $\mathbf{CLS}$ always outperforms $\mathbf{DLS}$ even with a single relay. It is important to observe that the performance of the $\mathbf{CLS}$ scheme reaches a peak for a higher number of sources. Also note that as the number of sources is increased, the performance of $\mathbf{DLS}$ deteriorates faster than the performance of $\mathbf{CLS}$ and this is only because the increased node density increases the interference. Another important conclusion from these results is that when the number of relays is equal to the number of sources, then the maximum performance can be reached. In general a higher number of maximum available relays increases performance since more options exist for the scheduling algorithm. However, the increased number of available relays cannot help when the node density is increased beyond a certain point (even when the number of relays is equal to the number of sources). For different $\beta_1=4$dB,$\beta_2=6$dB in Fig.~\ref{fig:results1}(b) the results present a similar trend while the peak performance has a minor increase. This result is very important and it actually means that significant variations in $\beta_1$ and $\beta_2$ lead to minor throughput differences. The reason is that for a lower $\beta_1$ the constraint on the cooperative link is tougher to meet through a higher $\beta_2$. This result with minor throughput variations was also obtained for even lower values of $\beta_1$ and supports our original choice of decoupling the two SINR constraints.

For a lighter traffic load of $B$=4 slots with $T$=8 in Fig.~\ref{fig:results2}(a,b), we see that the performance trend is similar. The throughput increase is now lower for a smaller number of sources due to the decreased traffic load. Also the peak performance does not reach the same level as before. Nevertheless, the cooperative relays still offer performance benefits in this case of lower traffic demand. The performance of our heuristic is also very good when compared with the the optimal solution, and is also consistent with the behavior of the randomized rounding approach~\cite{randomize-rounding}.

\section{Conclusions}
\label{section:conclusions}
In this paper we presented a MILP formulation for the problem of link scheduling in wireless networks that cooperate through an amplify-and-forward protocol. A simple MILP formulation is enabled by the separation of the SINR of the cooperative transmission into linear constraints that correspond to the direct and AF transmissions. %The practical advantage of our approach is that a non-linear constraint on the joint SINR expression would require a more complicated solution approach. Our experiments with different packet decoding thresholds indicate that the previous non-linear solution might not result in significant performance improvement.
The performance results indicate that as the number of sources is increased, it is more critical to employ the proposed approach that allows the scheduling algorithm to be more flexible in the allocation of individual source and relay transmissions across time.

\bibliography{../tony-bib}

\end{document}